\documentclass[12pt]{extarticle}

\usepackage[margin=1in]{geometry} 

\usepackage{stix}

\usepackage{amsmath}
\usepackage{amssymb}
\usepackage{amsthm}

\usepackage[colorlinks, citecolor=blue]{hyperref}
\usepackage[many]{tcolorbox}
\tcbuselibrary{theorems}
\tcbuselibrary{skins}
\tcbset{skin=enhanced}
\usepackage[explicit]{titlesec}
\usepackage{xcolor}
\usepackage[export]{adjustbox}
\usepackage{graphicx}
\usepackage{tikz}
\usepackage{microtype}
\usepackage{tabularx}

\usepackage{booktabs}

\newcolumntype{L}{>{\raggedright\arraybackslash}X}

\usetikzlibrary{external}

\definecolor{aqua}{HTML}{40DDDD}
\definecolor{crimson}{HTML}{DC143C}
\definecolor{pinkish}{HTML}{FA9DEE}
\definecolor{indigo}{HTML}{4B0082}

\newtcolorbox{evidence}{colback=blue!5, colframe=blue!60, boxrule=0pt, left=5pt, leftrule=5pt, enlarge top by=0.1cm, enlarge bottom by=0.1cm, breakable}
\newtcolorbox{directions}{colback=orange!5, colframe=orange!60, boxrule=0pt, left=5pt, leftrule=5pt, enlarge top by=0.1cm, enlarge bottom by=0.1cm, breakable}

\newtcolorbox{boldbox}{colback=yellow!10, colframe=yellow!60, boxrule=0pt, left=5pt, leftrule=5pt, enlarge top by=0.1cm, enlarge bottom by=0.1cm, breakable}
\newtcolorbox{infobox}{colframe=green!40, colback=aqua!6, boxrule=0pt, left=5pt, leftrule=3pt, breakable, enlarge top by=0.1cm, enlarge bottom by=0.1cm}

\makeatletter
\renewcommand\maketitle{
\begin{center}
{\Large \bfseries \@title}

\vspace{0.05in}

{Anthony Lin}

\vspace{0.05in}
\end{center}

}
\makeatother

\makeatletter
\renewcommand\paragraph{\@startsection{paragraph}{4}{\z@}
  {-3.25ex \@plus -1ex \@minus -.2ex}
  {-1em}
  {\normalfont\itshape}} 
\makeatother

\title{Attachment: a predictive coding approach}

\usepackage{apacite}
\usepackage[labelfont=bf]{caption}

\begin{document}

\maketitle

\begin{abstract}
	We introduce a novel predictive coding framework for studying attachment theory. Building off an established model of attachment, the dynamic-maturational model (DMM), as well as the neuroanatomical Embodied Predictive Interoception Coding (EPIC) model of interoception and emotion, we not only elucidate how neural processes can shape attachment strategies, but also explore how early attachment experiences can shape those processes in the first place.
\end{abstract}

\section{Introduction}
Attachment theory has long been part of the psychological canon, providing a compelling framework for understanding the development and importance of interpersonal bonds. Early attachments to caregivers have a significant influence on a countless range of developmental outcomes \cite<for a review, see>{Groh2017}, and it is through those attachments that individuals learn how to navigate the social world \cite{Atzil2018}.

Recent advances in computational neuroscience, including predictive coding, which models brain function in terms of Bayesian inference, as well as the Embodied Predictive Interoception Coding (EPIC) model \cite{Barrett2015, Barrett2017}, which models interoception and emotion as a predictive process and has promising implications for the treatment of conditions like depression \cite{Barrett2016}, provide an opportunity to elucidate the neural systems underlying attachment.

Under a predictive coding approach, the brain generates predictions about its sensations, and receives feedback from the discrepancy between its predictions and the actual sensations (prediction error). In response to prediction error, the brain can either update its predictions, or perform actions to make its predictions come true. As \citeA{Barrett2015} beautifully describes it, predictive coding sees the brain as ``an elegantly orchestrated self-fulfilling prophecy''.

In this paper, we bridge the gap between attachment theory and modern computational neuroscience by placing attachment in a predictive coding framework (specifically, the \textit{active inference} formulation of predictive coding described in \citeA{Parr2022}), building off the dynamic-maturational model \cite<DMM,>{CrittendenLandini2011}, an established model of attachment, as well as the EPIC model.

\section{Preliminaries}

\subsection{The EPIC model}

Predictive coding, formalized mathematically by \textit{active inference} \cite{Parr2022}, is an increasingly popular hypothesis which proposes that the brain is constantly sculpting an internal probabilistic model that generates \textit{predictions} of future sensations. It flips the typical feedforward approach on its head: instead of constructing representations from sensations, the brain proactively represents the sensations that it expects to observe. 

Whenever sensations fail to match predictions, they generate \textit{prediction errors} which the brain can resolve by either changing its model using Bayesian inference (perception), or performing an action which brings sensations closer to predictions (action). Predictive coding integrates control, learning, and uncertainty into a unified theory. 

The Embodied Predictive Interoception Coding (EPIC) model \cite{Barrett2015, Barrett2017} is a predictive coding theory of brain function at the systems level.\footnote{Before, the model had a life as the \textit{conceptual act theory} \cite{Barrett2006}, which proposed the provocative idea that emotions were not innate, but instead learned concepts which categorized patterns of interoceptive (body-based) and exteroceptive (environmental) sensations.} In the EPIC model, a concept is a prediction which categorizes sensations in a meaningful way \cite{Barrett2017}. The brain forms a predictive hierarchy where concepts can emerge at any level of perceptual processing, and higher-level concepts influence the probability of lower-level, subordinate concepts being activated.\footnote{For example, the semantic concept of ``tennis ball'' emerges in association cortices and strongly entails concepts of features like ``greenness'' and ``roundness'' in sensory cortices.} 

All the time, the brain -- or specifically, the default mode network \cite{Menon2023} -- prepares multiple mental simulations using its library of concepts, selecting the simulation and plan of action that is most likely to predict incoming sensations using Bayesian reasoning \cite{Barrett2017}. However, this selection process is biased towards more desirable outcomes as the brain sets a \textit{prior distribution} over predictions, making it more likely that we select the predictions which we prefer to make a reality \cite{Parr2022}.\footnote{That is why, say, if we are closer to starvation than fullness, our brain compels us to seek food instead of simply conceding to starvation. The prior probability of fullness is a lot greater than the prior probability of starvation.}

Cortical architecture implements a predictive hierarchy. At a circuit level, although there are cytoarchitectural differences between cortical areas, there is a canonical microcircuit \cite{Bastos2012} in the cortex where predictions and prediction errors flow through a hierarchy of cortical columns. In each column, layer III neurons encode a hidden state which the layer V neurons convert into a prediction signal that is sent to the column below, with layer IV neurons receiving prediction errors as feedback. Layer II neurons receive prediction signals from the column above which influence the activity of layer III neurons.

At a systems level, the limbic cortices, which consist of the posterior regions of the ventromedial prefrontal cortex (vmPFC) and orbitofrontal cortex (OFC), the anterior cingulate cortex (ACC), as well as the ventral regions of the anterior insula (aINS), are at the very top of the hierarchy \cite{Barrett2015}. They are agranular, which means they lack a defined layer IV, making them ill-equipped to receive prediction errors. Yet, they have extensive connections to the rest of the brain and play a key role in the visceromotor control of the autonomic, hormonal, and immune systems \cite{Kleckner2017}.

Concepts that originate in the limbic cortices are abstract, multimodal predictions that jointly categorize interoceptive and exteroceptive sensations, relating environmental states to bodily states \cite{Barrett2017}.\footnote{For example, the concept ``this pizza is tasty'' predicts the interoceptive feelings of pleasure associated with the exteroceptive sensations of seeing and eating pizza.} Concepts vie for expression through a competitive process in the cortex \cite<modulated by value signals in the OFC,>{Rolls2016} and the limbic basal ganglia (BG) which revolves around the nucleus accumbens (NAc) and ventral tegmental area (VTA) \cite{YinKnowlton2006}. 

Predictions from higher-level cortices place a prior distribution on the concepts that are activated in lower-level cortices -- for example, activating the concept ``banana'' would bias neurons in the primary visual cortex to expect a yellow object. Prediction errors are propagated in the reverse direction, updating the internal model at different levels of the hierarchy.

In particular, prediction signals from the limbic cortices cascade to lower-level cortices with more granular structure, as well as subcortical regions. Interoceptive predictions are dispatched to the mid-posterior insula and subcortical regions, which are also responsible for resolving prediction errors through visceromotor control \cite{Kleckner2017}. On the other hand, exteroceptive predictions travel through associative and sensorimotor regions of the cortex, turning the abstract concepts of the limbic cortices into concrete, embodied concepts \cite{Binder2011} which predict incoming sensations, with prediction errors either adjusting predictions or motivating movements.

\subsubsection{Precision and attention}
Since predictions and generated observations are both probability distributions, their \textit{precisions} (inverse variances) represent our confidence in those predictions and the reliability of those observations respectively \cite{Moran2013, Parr2022}. These precisions also influence the \textit{precision weighing} of prediction errors -- when we decrease the precision of predictions or increase the precision of their generated observations, we also increase the precision of their prediction errors. 

Precision weighing influences the sensations to which the brain attends, ensuring that the brain spends its substantial energy expenditure \cite{Raichle2010} on encoding and consolidating the most relevant signals \cite{Barrett2017}. Although cortical columns can implement precision weighing \cite{Bastos2012}, at a systems level, precision weighing is handled by the salience network. 

\subsection{The DMM}

The dynamic-maturational model (DMM) is a model of attachment that extends Mary Ainsworth's type A, B, and C classifications in the Strange Situation Procedure \cite<SSP,>{Ainsworth1978} in infancy to cover the entire lifespan. Developed by Patricia Crittenden, one of Ainsworth's graduate students, the DMM differs from other approaches to attachment by focusing on the self-protective functions of attachment over the nature of the attachment bond itself \cite{Crittenden2006}.

\begin{infobox}
The DMM is a result of many decades of research by Crittenden and her colleagues. For a comprehensive review of the DMM's history, see \citeA{Landa2013}. Crittenden's publication on the DMM \cite{Crittenden2006} gives an accessible overview of the DMM, while her book \textit{Raising Parents} \cite{Crittenden2016} provides a more detailed account.
\end{infobox}

The DMM originated from Crittenden's work in the 1980s on classifying aberrant patterns of attachment in the SSP. Some infants did not fit any of Ainsworth's three type A, B, or C classifications (which can be thought of as attachment \textit{strategies}), but instead showed a confused mix of behaviors which involved a mix of type A and C behaviors, disorientation, and threat responses \cite{LandaDuschinsky2013}.

Another one of Ainsworth's graduate students, Mary Main, was also working on resolving this exact same paradox, proposing that those behaviors simply represent ``disorganized'' attachment patterns representing the failure to organize a successful attachment strategy \cite{Main1981}. 

However, Crittenden had a different view. Focusing her research on severely maltreated infants, she saw those ``disorganized'' patterns reflected not maladaptation, but instead the wondrous ability for infants to adapt even to the most dangerous situations. In her PhD thesis \cite{Crittenden1983}, she proposed that the aberrant behaviors were actually strategic combinations of the type A and C strategies (a type A/C strategy), a clever way to navigate the approach-avoidance conflict with caregivers who created it.

Instead of seeing attachment strategies merely as a way to maintain proximity to a convenient caregiver as Main had, Crittenden saw attachment strategies as a way to maintain the \textit{availability} of a caregiver for protection  \cite{Landa2013}. And this is a crucial difference that distinguishes Crittenden's thinking from other researchers in attachment.

Attachment research has traditionally been focused on the nature of the attachment bond itself, measuring the degree to which a child avoids or worries about their attachments with caregivers. Yet, Crittenden first starts from the maxim that every child \textit{must} maintain the availability of their caregivers for protection, and then examines how each attachment strategy allows the child to do so even in the face of possible adversity \cite{Crittenden2021}. Type A strategies allow children to maintain attachments to maltreating but predictable caregivers, while type C strategies allow children to maximize the availability of caregivers who give unpredictable signals about their availability \cite{CrittendenLandini2011}. 

Part of Crittenden's reasoning for rejecting disorganization stems from her maxim that children must maintain caregiver availability at all costs \cite{LandaDuschinsky2013} -- after all, if an attachment strategy had truly failed to protect the child, then they would simply be unable to survive, let alone have some sort of attachment to their caregivers. It is very much congruent with John Bowlby's ethological approach that places attachment in the context of a grander struggle for survival \cite{Bowlby1969}.

Furthermore, Bowlby had proposed that, when attachment figures are unavailable, an individual can omit information from awareness to protect themselves from the pain of unavailability, in what is called \textit{defensive exclusion} \cite{Bowlby1980}. Building upon that idea, Crittenden proposed that, during processing, type A strategies involve omitting information about internal feelings (affect), and type C strategies involve omitting information about external contingencies (cognition) \cite{Crittenden2000}.

Crittenden has since built upon her earlier work to create the DMM \cite{Crittenden2006} as a culmination of all her ideas. The DMM expands the two type A and type C strategies (A1-2, C1-2) in the SSP \cite{Ainsworth1978} into eight strategies for each category (A1-8, C1-8). The added strategies (A3-8, C3-8) are only made possible through maturation beyond infancy. In addition, the four type B strategies (B1-4) in the SSP are expanded to five strategies (B1-5). The type A and C strategies can be combined not only in alternating way (type A/C) depending on the context, but also used simultaneously (type AC) as well.

\section{Framework}
\label{sec:allo}

Our framework is based on a fundamental organizing principle of the brain: it is constantly anticipating our future resource needs so it can prepare to satisfy them before they arise \cite{Sterling2015, Barrett2017}. This process is called \textit{allostasis} \cite{Sterling2012}, and the ability to plan around future needs is one of the chief advantages of a predictive brain. Predictions can extend into arbitrarily long timescales, incorporating our best guess about the resource needs that will emerge in the future and the ways by which we can fulfill those needs. Our ultimate goal is not only to maintain stability, but to maintain stability \textit{through} change \cite{McEwen2000}, and that means anticipating those changes so that we can prepare for them ahead of time.

From an allostatic lens, attachment is simply another resource that can be recruited in response to danger \cite{Atzil2018, Long2020}. We can seek attachment figures not only to request their support, but also to obtain evidence that they are available in the first place even in the absence of a present threat.

For infants, allostasis almost entirely revolves around attachment since they are unable to manage their needs on their own: they anticipate their resource needs, and they recruit as much care from their caregivers as they need to meet those needs. With maturation, they gradually learn to meet more of their needs independently. But even in adulthood, attachment still plays a key role in allostasis, since when we lack support from others, we have to spend additional energy to ensure our needs are fulfilled \cite{Coan2015}. Much has been written about the minutiae of attachment bonds themselves, such as the stance of an individual towards their attachment figures, but the only reason those details matter is due to their contribution towards allostasis.\footnote{Without keeping this in mind, it is easy to misinterpret attachment behavior. For example, seemingly stoic responses to separation (A1-2 strategies in the DMM) have been interpreted as a ``detachment'' from caregivers \cite{Bowlby1969} or ``avoidant'' behavior \cite{Ainsworth1978}, but that betrays the possibility that it could be a way of maintaining attachment to caregivers who punish expressed negative affect \cite{Crittenden2016}.}

As such, our framework centers around the following three \textit{stages}, outlining a cycle by which we adapt to current and future danger:

\begin{itemize}
	\item \textbf{Stress:} Prediction errors represent a potential failure of allostasis. The stress response increases our level of arousal in response to a prediction error that cannot be resolved with any available concepts or actions.
	\item \textbf{Seeking:} During stress, we may seek the availability or support of attachment figures who can help us resolve the prediction error. The manner in which we do so depends on the magnitude of the prediction error and the traits of our attachment figures.
	\item \textbf{Resolution:} We resolve the prediction error and terminate the stress response, which may or may not involve the help of attachment figures. We encode and consolidate the prediction error as a new concept which helps us cope with similar situations in the future.
\end{itemize}

\subsection{Stress}
Allostasis is \textit{predictive regulation} \cite{Sterling2012}, so ineffective allostasis is either a failure in prediction (\textit{ambiguity}, a failure to accurately anticipate sensations) or in regulation (\textit{risk}, a failure to maintain an adequate state). Since prediction errors can either result from ambiguity or risk \cite{Parr2022}, they are the currency of allostatic failure, a debt which the brain must pay off.

 Ideally, any prediction errors can be quickly resolved by selecting a concept which better predicts our sensations (changing our prediction), or by changing our plan of action (changing our regulation). But when there are no suitable concepts or policies, we start to experience \textit{stress}, a response that initiates a cascade of arousing neural and metabolic effects \cite{McEwen2015} which aim to resolve the prediction error \cite{Peters2017}. 
 

\subsubsection{Mechanisms}
 
Stress leads to a wide range of effects on the brain and body, but we will only focus on two of these effects, both of which are particularly important for stimulating attention and learning during stress.

First, the locus coeruleus (LC) releases norepinephrine (NE) both tonically and phasically \cite{AstonJones2005, Valentino2008}, increasing energy expenditure in the central nervous system \cite{ODonnell2012}. Tonic LC activity promotes general vigilance, while phasic LC activity selectively increases the precision of ascending precision errors \cite{Mather2016, Peters2017}. In general, LC activity reflects the magnitude of unexpected prediction errors and increases the rate of learning in preparation \cite{Jordan2024}.

Second, the HPA axis, which is primarily activated by projections to the paraventricular nucleus of the hypothalamus (PVN) \cite{UlrichLai2009}, releases cortisol, which binds to mineralocorticoid and glucocorticoid receptors in the cortex and amygdala, modulating synaptic plasticity in those regions \cite{Schmidt2013, Myers2014}. Usually, acute stress enhances long-term depression (LTD) and impairs long-term potentiation (LTP). LTD destabilizes the internal model, paving way for the encoding of new concepts through LTP once the prediction error is resolved \cite{Peters2017}.\footnote{In some cases, stress can disrupt memory reconsolidation \cite{Akirav2013}, which facilitates the permanent forgetting of concepts that are no longer relevant \cite{Nader2009}.}

The LC and PVN are both well-positioned to integrate signals across the brain to generate stress responses. The LC receives inputs from almost every brain region \cite{Schwarz2015, BretonProvencher2019}, and while the PVN predominantly receives inputs from subcortical regions, cortical regions can indirectly influence its activity via the medial and central nuclei of the amygdala (MeA and CeA), the bed nucleus of the stria terminalis (BNST), the nucleus of the solitary tract (NTS), and a number of other hypothalamic regions \cite{Herman2003}. These responses are separate from each other \cite{UlrichLai2009}, with the LC providing an immediate noradrenergic response which regulates attention and energy expenditure, and the PVN providing a much slower HPA axis response which further supports its longer-term role in modulating synaptic plasticity.

Stress is essentially a metabolic investment by the brain. When the investment bears fruit, the brain not only resolves the prediction error, but potentially also learns new concepts so that similar situations are more predictable in the future. But not all investments are made equal, and chronic stress has a litany of effects on the brain and body \cite{Lupien2009}.

\subsubsection{The need to learn}

Traditionally, stress is regarded as a response to \textit{risky} situations, where we know the threat, we know the likelihood of it happening, and we want to do something about it. Yet, risk is not the only source of prediction error and not the only source of stress. Stress can also be a response to \textit{ambiguous} situations where our internal model is unable to make accurate predictions.

It is useful to see concept learning from an allostatic lens where concepts are just like any other resource. Every situation demands that we have the right array of concepts to successfully predict incoming sensations and minimize ambiguity. Without those concepts, we experience an influx of prediction error, and if the prediction error is too high, we are compelled to avoid the situation entirely since there is too much potential for unexpected threat. So every time we learn a new concept, we are actually doing so in anticipation of the conceptual demands of future situations. 

From this lens, allostasis is not just about energy regulation, but also about ambiguity regulation, ensuring that future situations are not surprising to the point of danger. The brain exacts a high metabolic cost \cite{Raichle2010}, so it is preferable for the most frequently occurring concepts to be learned as early as possible when the brain is most malleable to new learning. It has even been suggested that the reason childhood growth is slower in humans than in other primates is to allocate more energy to the brain during development \cite{Kuzawa2014}. 

Attachment relationships are a \textit{secure base} for exploration and learning \cite{Bowlby1988}. When a child's short-term resource needs are buffered by caregivers, they are free to focus on obtaining the resource that will eventually earn them more independence: concepts. Children learn to be independent not by being left alone, but by having the security of a dependable caregiver. It is dependence that buys independence.

Furthermore, stress is no longer merely a defensive response, because the brain may also deploy stress responses in anticipation of concepts that will need to be learned in the future. We could even understand rumination as the ongoing synthesis of existing knowledge to form new concepts in anticipation of future events. Rumination is a known function of the default mode network \cite{Zhou2020, Menon2023}, which aligns with its general role of anticipating the body's resource needs in support of allostasis \cite{Barrett2017}.

There are multiple lines of evidence suggesting that reducing ambiguity is a major imperative of the brain. This includes ambiguity aversion in behavioral economics research \cite{CamererWeber1992}, neuroimaging studies implicating the frontopolar cortex (FPC) and dlPFC in mediating exploratory behavior \cite{Daw2006, Badre2012, Hogeveen2022, Tomoda2011} \cite<including causal evidence that inhibiting those regions can abolish certain exploratory behaviors,>{Zajkowski2017, Toghi2024}, amygdala responses to novelty \cite{Weierich2010, Blackford2010, Balderston2011, Balderston2013}, and phasic NE responses to any uncertainty that cannot be explained by the internal model \cite{Yu2005, Dayan2006}.

\subsection*{Seeking}

Children are born without the ability to do all but the most primitive actions.  Endowed with barely any knowledge beyond the scant instructions encoded in their genome, they are not only tasked with staying alive at the most vulnerable stage of their life, but they also face the monumental challenge of spending their first two decades on Earth learning the concepts necessary to survive independently in a world that has far outpaced our Pleistocene-era arboreal niche. These are two huge allostatic demands: they have to manage their physiological needs with barely any ability to do so alone, while also learning a myriad of concepts without compromising their safety. Luckily, evolution has given the child a wonderfully simple answer in the form of attachment, reducing the grand challenge of allostasis to the problem of seeking the availability of their caregivers.

In early childhood, the default response to stress is seeking attachment \cite{Bowlby1969}, and other responses are learned only through maturation. Furthermore, children form attachments to caregivers regardless of caregiving quality. This is supported by numerous studies using animal models \cite{Sullivan2012}, including studies on rodents \cite{Moriceau2006, Raineki2010, Sullivan2000} and nonhuman primates \cite{McCormack2009, Roma2007}.\footnote{In rodents, the maternal odor lowers corticosterone levels and blocks fear learning, while NE, which is abundant in the infant brain, encodes a learned odor preference, so that an attachment to the mother can be learned regardless of caregiving quality \cite{Perry2014}.}. Furthermore, Crittenden's early research demonstrates that preschool-aged children develop strategies to maintain attachment to maltreating caregivers, including taking care of caregivers in an act of role-reversal \cite<the DMM A3 strategy,>{Crittenden1992}, as well as compulsively complying with abusive caregivers \cite<the DMM A4 strategy,>{Crittenden1988}.

\subsubsection{Attachment concepts}

Much has been said about \textit{internal working models} of attachment, although their exact representation in the brain remains a question \cite{Bretherton2008}. However, we propose that, instead of being represented by a separate attachment module, these models \textit{emerge} from the basic predictive architecture of the brain, forming when exteroceptive cues are associated with interoceptive cues representing the outcomes of interactions with attachment figures \cite<for a review of evidence, see>{Atzil2018}. 

Now, this does not mean that there is no biological basis for attachment. Indeed, certain adaptations in our recent evolutionary history support the formation of long-term attachments, including the elaboration of neurotransmitter systems and the expansion of the brain \cite{Feldman2017}. It is just that these adaptations merely support the predictive architecture in navigating and learning from the social world, and do not define a separate attachment module with fixed percepts and strategies. In general, there is no ``social brain'', but rather a brain that can learn to be social \cite{Heyes2015}. Attachment must be learned through experiences which demonstrate that attachment behaviors are relevant to allostasis \cite{Atzil2018}.

In our view, every attachment figure is represented by \textit{attachment concepts} instantiated in prefrontal and multimodal association cortices, storing the outcomes of different interactions with them in different contexts. Attachment concepts can pertain to the attachment figure's \textit{availability} (whether they will respond to distress) and \textit{support} (which stressors they can help resolve). Even if there are biological mechanisms which facilitate their formation and selection, attachment concepts still compete with other high-level concepts for expression in default mode network simulations \cite{Barrett2017}.

\paragraph{Construction} Attachment concepts are distributed across the brain \cite{Feldman2017}, since social behaviors reuse the predictive architecture of the brain and integrate inputs from many different modalities. Similar to the proposal of \citeA{Binney2020} that frames social cognition as a special case of semantic cognition, we propose that attachment concepts are embodied, multimodal semantic memories \cite{Binder2011} constructed by the same regions responsible for other semantic memories. Even if the idea of an ``internal working model'' of attachment implies the existence of a special attachment module, attachment concepts are simply semantic memories like any other. As we will explore later, this means that \textit{semantic representation} is key to healthy attachment. As decisive as early attachment behaviors might be, even individuals who have faced adversity can, with maturation, begin to reflect on their experiences, reorganize their memories, and achieve lasting change \cite{Crittenden2016}.

Individuals are represented not as a single concept, but as a range of concepts which depend on their mental states and other contextual details (including our own mental states). \textit{Mentalization} is precisely the ability to infer the mental states of the self and others \cite{Fonagy1997}, allowing us to select the most applicable concept to represent an individual in a given context. Mentalization is crucial for attachment, especially after early childhood where children must cope with longer separations from caregivers (making it more important to anticipate their availability) and face the challenge of building reciprocal peer relationships.

The anterior temporal lobe (ATL) acts as a hub which constructs multimodal concepts by integrating subordinate concepts from different modalities \cite{Rice2015, Binney2016}. It is no surprise that the ATL is essential for representing semantic information about other individuals, such as their identity and personal significance \cite{Olson2013}, since this information is inherently multimodal. The representation of concepts which categorize mental states (such as traits and emotions) additionally relies on the medial prefrontal cortex (mPFC) \cite{Wagner2012} which encodes a common neural code for each concept \cite{Skerry2014, Chikazoe2014, Weaverdyck2021}, lining up with the region's general purpose of representing abstract schemas. Mental states are assembled from these mental state concepts \cite{Lin2023}, and there is even preliminary evidence \cite{Thornton2019} that the brain represents individuals in part as an amalgamation of their perceived traits. 

Yet, human cognition is unique not because of our ability to represent abstract concepts, but because of our ability to adapt those concepts in a context-sensitive way (\textit{semantic control}) \cite{LambonRalph2017}. We can think of mentalization as simply an application of semantic control, adapting a concept of an individual to the present context. Semantic control involves a network consisting of the dorsomedial prefrontal cortex (dmPFC), supplementary motor area (SMA), pre-supplementary motor area (pre-SMA), inferior frontal gyrus (IFG), posterior middle temporal gyrus (pMTG), and posterior inferior temporal gyrus (pITG) \cite{Jackson2021}.\footnote{The temporoparietal junction (TPJ) has also been implicated in mentalization tasks \cite{Frith2003, VanOverwalle2009}, but we propose that the TPJ mainly plays a role in updating mental states instead of representing them, based on the \textit{contextual updating} theory \cite{Geng2013} which implicates the TPJ in responding to changes in context and recruiting updates to the internal model accordingly.} In addition, the hippocampus is vital for contextual processing, binding representations across cortical regions without the need for direct long-range cortical connections \cite{Rolls2016, Yonelinas2019}.

Finally, it is vital to be able to link attachment concepts to their allostatic value. The insula processes interoceptive signals about the state of the body \cite{Kleckner2017}, while the OFC encodes the value of concepts \cite{Rolls2016} and the vmPFC constructs a cognitive map of choices (and their outcomes) based on their value \cite{Veselic2025}. The BG also plays a key role in biasing the selection of concepts. Concepts representing outcomes compete in the limbic BG (targeting the ventral striatum, particularly the NAc), while concepts representing actions compete in the associative and sensorimotor BG (targeting the dorsal striatum). Both striatal regions receive dopaminergic modulation, with the VTA modulating the ventral striatum and the substantia nigra pars compacta (SNc) modulating the dorsal striatum \cite{YinKnowlton2006}.

\paragraph{Maturation} We propose that attachment concepts rapidly evolve in complexity over the first 2 years of life, where the default mode network goes from being barely developed to taking on its adult-like topology \cite{Fransson2007, Fransson2011, Gao2009}. Attachment concepts are not just descriptions of others -- they are manifested through \textit{embodied}, context-dependent simulations of social interactions initiated by the default mode network \cite{Barrett2017, Atzil2018}. As more sophisticated mental simulations become available, individuals can adapt to more challenging relational contexts. This does not mean that the attachment system is mature at 2 years of age -- the default mode network continues to mature all the way up to adulthood \cite{Rebello2018} and the child still needs to learn a myriad of attachment concepts -- but much of the groundwork for the attachment system is already in place at that point.

At birth, the default mode network is barely developed, with traces only found in sensorimotor regions \cite{Fransson2007, Fransson2011}. During the first 6 months of life, functional connectivity between the posterior cingulate cortex (PCC), inferior parietal lobule (IPL), and hippocampal formation (HF) rapidly increases \cite{Gao2015}. PCC-HF interaction is involved in retrieving declarative (semantic and episodic) memories for mental simulations \cite{Leech2014, Rolls2019}. We propose that this is related to the finding that, at 6 months of age, children begin to exhibit basic attachment behaviors like proximity-seeking \cite{Bowlby1969}. These behaviors show that the child recognizes the caregiver as relevant to allostasis even in the absence of an immediate stressor, which requires the ability to encode and retrieve semantic, multimodal representations \cite{Binder2011} of the caregiver as well as the episodic memories of their caregiver's behaviors (particularly those of their needs being met) over time \cite{Lehn2009}.

From 6 months of age, functional connectivity between the two core hubs of the default mode network, the PCC and the mPFC, starts to rapidly increase before reaching robust coupling by 1 year of age \cite{Gao2015}. Why might this matter for attachment? Consider that the mPFC is essential for storing \textit{causal} associations between actions and outcomes \cite{Morris2022}. It is plausible that the mPFC could support the simulation of outcomes contingent on an individual's own actions, which is a vital part of forming the self-model that has been ascribed to the default mode network \cite{Levorsen2023, Menon2023}. This is further supported by the finding that children can only \textit{act} upon action-outcome contingencies starting at 12 to 15 months of age \cite{Elsner2003} although they are aware of those contingencies earlier \cite{Baumgartner2011}. The mPFC also stores \textit{schemas}, or abstract concepts which generalize across episodes \cite{Sekeres2018, Audrain2022}, allowing for abstract, multimodal predictions of availability and support that are not anchored in highly specific cues. Altogether, not only can the child assemble attachment behaviors into an organized strategy -- aligned with the fact that the SSP is only applicable to children 11 months of age or older \cite{Ainsworth1978} -- but they can also also begin to learn voluntary actions which increase the availability of their caregivers. The latter is characteristic of the A3-4 and C3-4 strategies in the DMM which are both more strategic than simply modulating the level of expressed distress as in the A1-2 and C1-2 strategies \cite{CrittendenLandini2011}. For example, children who have suffered abuse can learn to compulsively comply with caregiver demands (A4) starting around 18 months of age \cite{Crittenden1988, Crittenden1992}.

Between 1 and 2 years of age, there is limited change in default mode network activity other than its increased coupling with the parahippocampal cortex (PHC), and at 2 years of age, the default mode network finally reaches an adult-like topology \cite{Gao2009}. The PHC is involved in processing \textit{contextual associations}, or processing associations between concepts in a context-dependent way \cite{Aminoff2013}, which is vital for making predictions at the level of \textit{events} and not just individual stimuli.

Combined, these three milestones facilitate the development of mentalization. Generally, children start to exhibit mentalization abilities around 18 months of age, although explicit mentalization emerges only around 4 to 6 years of age \cite{Frith2003}. To mentalize, the child must predict the outcomes of their own behaviors and others' behaviors, all in the context of latent mental states which can only be inferred based on the past history of interactions as well as other contextual details. This requires all of the capabilities mentioned above: declarative memory, action-outcome learning, and contextual associations. Indeed, mentalization tasks consistently activate default mode network regions \cite{Spreng2009}, and the default mode network has even been called the ``mentalization network'' \cite{Atzil2018} even though mentalization is not its only function.

\subsubsection{Planning}

It has been proposed that the goal of attachment behaviors is to seek the proximity \cite{Ainsworth1972} or availability \cite<as in the DMM,>{LandaDuschinsky2013} of attachment figures, but here, we take a more nuanced approach.

Ultimately, attachment behaviors are a means to an end, ensuring attachment figures are available to provide the support needed to resolve stressors. While we agree with Crittenden that maintaining availability is a key goal, the chief motivation for maintaining availability is due to the knowledge that attachment figures can provide support that is relevant to predicted stressors. Furthermore, support is multi-dimensional: some attachment figures might be better at helping resolve a certain stressor than others. For example, some individuals might be better at providing material support, while others might be better at providing emotional support.

We propose that, during stress, the brain plans attachment behaviors \textit{depending on} the prediction errors that need to be resolved, which may or may not be directly related to attachment needs. The default mode network interacts with the partially overlapping semantic network to retrieve attachment concepts that are most relevant to resolving the prediction error -- through both direct cortical connections \cite{Peters2013, Rolls2016} as well as indirectly through the BG (through both dopaminergic modulation and nigrothalamic projections) \cite{Yin2023} -- while the salience network and LC increase sensory precision across the brain to facilitate concept selection. During mental simulation, the brain predicts the present and future mental states \cite{Thornton2019} of attachment figures, and uses those mental state predictions to infer which interactions are possible with them \cite{Thornton2024} and whether they are available to provide the needed support. Eventually, the brain devises a plan which may involve both seeking the attachment figure to ensure their availability, and calling for their support as necessary. This process might not even involve any observable behavior if the brain is sufficiently confident in its predictions, which is more common in balanced (type B) attachments.

\subsection{Resolution}

\subsubsection{Learning}

Once a prediction error is resolved, any new concepts that were acquired are encoded in an \textit{engram}, or a unique population of neurons responsible for storing a memory. Existing engrams may also be strengthened as well. Engrams are distributed across multiple brain regions \cite{Roy2022}, and often involve the hippocampus \cite{Sekeres2018} which can bind representations in different cortical regions together \cite{Rolls2016, Yonelinas2019}.

However, engrams are initially fragile, and they are easily forgotten without further \textit{consolidation}, which refers to the processes that transform an engram into a long-term memory. Consolidation relies on gene expression and protein synthesis, and reactivating a concept that was recently acquired (either by direct exposure or by imagination) drives the consolidation and strengthening of its engram \cite{Guskjolen2023}. It is one of the ways that rumination can either be adaptive, when it reinforces concepts that will be important in the future, or maladaptive, when it strengthens irrelevant concepts or transforms them in an unproductive way.

\subsubsection{Habituation}

Sometimes, stress is not worth the investment, and it is better to cease the stress response than to risk lasting damage to the brain and body \cite{Godoy2018}. In those cases, the brain may \textit{habituate} to the prediction error by abolishing its belief that it can achieve the goal state \cite{Peters2017}. During habituation, the brain chooses a more achievable goal state and suppresses any residual prediction error. It is essentially accepting that the prediction error is outside of its control.

Traditionally, habituation is defined as a decrease in response to a benign, repeatedly encountered stimulus. However, we extend the definition of habituation to include any process that resolves a prediction error by reducing its precision \cite{Ramaswami2014}. Instead of epistemic uncertainty that demands learning, the brain treats the prediction error as simply yet another random event to which it does not have to attend (aleatoric uncertainty). As a result, habituation can (and does) occur with events that are not benign, most importantly when habituating to stress \cite{Grissom2009}. 

Habituation immediately alleviates stress by resolving prediction error in the short-term \cite{Peters2017}. Yet, if it means abolishing a desirable goal state, habituation is essentially an allostatic bankruptcy, a maneuver which forgives the debt of prediction error -- but at a cost. Not only does it mean accepting additional risk, but also forfeiting the possibility of learning new concepts from the prediction error.

However, couldn't we simply habituate to every sensation to resolve prediction errors for good? This question is a variant of the ``dark room problem'' in active inference \cite{Friston2012}, and the answer is simple: the predictions we can make are constrained by their desirability. Limbic cortices are biased towards more desirable predictions, and since they lack a defined layer IV which would allow them to receive prediction error signals \cite{Barrett2015}, lower-level areas are disposed towards implementing the predictions imposed by the limbic cortices even amid prediction error. We can only habituate to a point before we are forced to extricate ourselves from the situation entirely as the risk becomes too high.

\subsubsection{Early experiences shape attention}

Both Bowlby \cite{Duschinsky2020} and Crittenden \cite{Crittenden2021} are bold in their supposition that early attachment experiences have a major influence on future pathology. Crittenden has proposed that attachment is a chief contributor to a diverse range of conditions such as eating disorders \cite{Ringer2007}, neurodevelopmental disorders \cite{Crittenden2016}, and personality disorders \cite{Crittenden2011}.

Of course, attachment is not everything, and many other factors (such as socioeconomic status) play a role. However, there are multiple lines of evidence which suggest that early experiences, which predominantly revolve around attachment, profoundly shape the development of \textit{attention} -- that is, the process of deciding which prediction errors should be prioritized for processing and encoding \cite{Feldman2010}. Attachment issues in early childhood can lead to biases in attention that are maladaptive later in life, preventing important concepts from being learned and jeopardizing allostasis.

To give some background, during early development, cortical processing gradually transitions from bottom-up feedforward processing, which shapes cortical structure and function, to top-down predictive processing. Initially, local connections are dominant in the cortex, with long-range connections (required to rapidly carry top-down signals directly to sensory cortices) gradually developing in the first few years of life \cite{Gao2011}. Myelination is most rapid in the first year of life, proceeding in a posterior-to-anterior direction so that sensorimotor cortices are prioritized for myelination before association cortices which integrate sensory inputs \cite{Tau2010}. Cortical hubs like the salience and default mode networks, which inherently require top-down processing to function, are underdeveloped in infancy \cite{Fransson2007, Fransson2011} with traces found only in sensorimotor regions. There is also evidence from primate studies \cite<for a review, see>{Price2006} that the pruning of corticocortical connections is only common for feedback projections and not feedforward projections, indicating that the structure of the predictive hierarchy develops over time. Noradrenergic LC neurons, which facilitate learning amid ambiguity \cite{Yu2005, Dayan2006}, are hyperactive in the infant brain \cite{Saboory2020}. Altogether, this means that infancy is an absolutely crucial period for shaping attention and future learning. Since infants are almost entirely dependent on caregivers in maintaining allostasis, they prioritize encoding the prediction errors which their caregivers help them to resolve, and thus are relevant to allostasis in the short-term \cite{Atzil2018}. 

Furthermore, there are a number of \textit{sensitive periods} in brain development, where there is heightened plasticity in neural circuits supporting a specific skill \cite{Knudsen2004}. Sensitive periods have been found for all five sensory systems \cite{Lewis2005, Kral2013, Ardiel2010, Schaal2020, Ventura2013} and possibly even for interoceptive functions \cite{Murphy2017}. Sensitive periods bias attention by their influence on the concepts that are learned.

Finally, \citeA{Teicher2016} reviews the extensive evidence that details the specific effects of childhood maltreatment on the brain, concluding that maltreatment enhances attention to threat as shown by differences in neural response within the dACC and amygdala (precision weighing and stress), hippocampus (context), as well as the vmPFC (fear extinction).

Altogether, these findings demonstrate that early attachment experiences are crucial for shaping the development of attention. Initially, bottom-up processing is dominant, and the youngest infants attend to stimuli in a relatively unbiased manner. However, as more and more concepts are learned, this shifts to a more self-fulfilling, top-down mode of processing. We are no longer looking to learn new concepts as much as we are looking for evidence to support our existing predictions.

\section{Applications}
\label{sec:app}
As a brief application of our framework, we will propose new hypotheses about the type A and C strategies described by the DMM \cite{CrittendenLandini2011, Crittenden2016}. Unless otherwise mentioned, every claim in this section is merely a hypothesis entailed by our framework which would require further evidence to confirm.

\subsection*{Type A}

Type A strategies arise as a result of consistent neglect or abuse in development, and involve inhibiting negative affect, ignoring the negative characteristics of attachment figures, and blaming the self for issues in relationships \cite{CrittendenLandini2011}. 

In infancy, type A strategies manifest as the inhibition of negative affect (A1-2) \cite{Crittenden2016}, as shown by a muted response to separation in the SSP \cite{Ainsworth1978}. This is a result of habituating to the stress of having their needs unmet. Since they are unable to meet their needs on their own as infants -- that is, they can only control their interoceptive state with the help of their caregivers -- habituation is the only choice in consistently neglectful or abusive environments. In the short-term, this alleviates stress and allows them to conserve resources that would have been directed towards a futile stress response.

Yet, constant habituation decreases interoceptive attention, blunting the precision of interoceptive prediction errors over time. Without accurate interoception, type A individuals struggle to appraise situations, since it is precisely those interoceptive signals which provide direct evidence of safety and threat.

Children learn to navigate the social world by forming associations between exteroceptive and interoceptive sensations during attachment experiences, and they can draw upon these associations to quickly appraise their safety in social contexts through mental simulation. Unfortunately, type A individuals face difficulties \textit{integrating} interoception into attachment concepts \cite{Crittenden2006} since they lack access to interoceptive information. Their brain might even deprioritize the formation of attachment concepts, since their caregivers are frequently unsupportive (and irrelevant to allostasis). Indeed, in AAI transcripts, type A individuals tend to give very brief responses that are suggestive of their impoverished attachment concepts, and struggle to describe their interoceptive states, omitting affective words and focusing on exteroceptive aspects (like action contingencies) instead \cite{CrittendenLandini2011}. Also, without access to interoception, type A individuals might struggle to mentalize since they lack suitable mental state representations -- for example, they might not understand the importance of supporting someone else in distress since they have seldom been supported when they were in distress themselves.

In other attachment models, type A strategies are often associated with being avoidant \cite{Fraley2000, Griffin1994} or dismissing \cite{Hesse2008} of close relationships. However, this is not necessarily due to a fear of intimacy. Close relationships \textit{demand} the ability to quickly appraise safety -- and since their attachment concepts lack interoceptive information, type A individuals struggle to make those appraisals.

Once a type A individual has reached preschool age, they may learn to recognize cues that anticipate caregiver maltreatment, and they may adopt a compulsively caregiving (A3) or compliant (A4) strategy to preempt that maltreatment \cite{Crittenden1988, Crittenden1992, Crittenden2016}. Prediction errors which suggest the threat of maltreatment suddenly become very salient. As a result, their attention is heavily biased towards stimuli that either confirm or deny the possibility of maltreatment, and they might become adept at mentalizing which of the caregiver's mental states might suggest maltreatment. Yet, although these strategies give them a way to appraise and create safety in their attachments with maltreating caregivers \cite{CrittendenLandini2011}, they do not restore access to interoception (except for the prediction errors associated with maltreatment).

During the adolescent years, the natural pressure to socialize and enter into relationships can be tremendously stressful, and this is when the lack of suitable attachment concepts especially starts to rear its head for type A individuals.

Relationships, whether platonic or sexual, require a lot of knowledge to successfully navigate. Even the meaning of relational categories like friendship, which many take for granted as obvious, is learned in childhood through early social experiences. Under typical development, while younger children view friendship primarily as involving proximity and shared activities, older children gradually learn to recognize the value of mutual support, loyalty, and trust in friendships \cite{Furman1983, Bigelow1977, Reisman1978}, which are all attachment concepts which involve mentalization as well as long-term safety appraisals, which both require accurate interoception to learn. Adolescents' concepts of romantic relationships share moderate overlap with their concepts of caregiver and friendship attachments \cite{Furman2002, Furman2008}. 

When that core social knowledge is missing in the confusing milieu of adolescence, serious problems are ought to ensue. Type A adolescents might not only struggle to build close friendships, but also struggle in romantic relationships \cite{Crittenden2016}, which require the learning of yet another relational category involving a plethora of new concepts \cite{Simon1984}. Concepts of intimacy are likely to be arcane to type A adolescents whose experience of attachment had revolved around surviving neglect and abuse. It is during adolescence that the compounding debt of conceptual poverty starts to seriously hurt. It also does not help that their internal model is geared towards expecting maltreatment from others. 

As a result, in long-term romantic or even platonic relationships, type A individuals are driven towards avoidance due to the uncertainty that comes with intimacy. It is not even necessarily because they \textit{fear} intimacy itself, but simply because intimacy is unfamiliar for them, creating too much prediction error to resolve at once. Romantic relationships can especially be a problem since physical touch produces powerful interoceptive sensations, and for type A individuals who have suppressed interoceptive prediction errors for so long, that creates a deluge of unexplained sensations which can make them feel endangered.

To avoid the stress and uncertainty of a long-term relationship while still obtaining some comfort from sexual activity, some type A individuals enter into a string of short-term sexual relationships (A5). Unlike long-term relationships (romantic or even platonic) which require an array of attachment concepts, short-term sexual relationships progress according to telegraphed scripts with clear contingencies \cite{MatickaTyndale1997, Paul2002, Luz2023}, which are a lot easier to learn and do not require much prerequisite knowledge about intimacy. Even then, the idea of sexual activity is frightening for some type A individuals, who settle for making a lot of acquaintances instead (A5-). As odd as it is to say, short-term relationships might be the only relationships that are within their zone of proximal development. 

However, we propose that A5 (including its platonic version, A5-) might be better described as ``compulsively scripted'' strategies rather than as ``compulsively promiscuous'' strategies. To cope with an underdeveloped attachment concept system, A5 individuals seek to learn telegraphed social scripts which allow them to seek interaction without much demand for intimacy or mentalization. Aside from short-term sexual relationships, groups with  rigid social rules and clear group identity markers, such as many niche subcultures and cults, offer a scaffolded milieu for A5 individuals, by allowing them to maintain good social standing without mentalization (as they can simply follow the rules), and by making it easier to identify group members with whom group-related topics are safe to discuss. This is related to the notion of \textit{tight cultures} at the national level, where nations that have faced severe historical threats tend to have stronger norms and a lower tolerance for deviant behavior \cite{Gelfand2011}. Furthermore, settings that allow for participation in a common activity without intimate interaction might be appealing to A5 individuals. For example, club and rave settings not only provide a common activity to do with others (dancing) with a superficial amount of interaction, but the music also offers a  predictable stimulus which entrains neural activity in regions responsible for affect \cite{Trost2014, Vuilleumier2015, Trost2024}, protecting A5 individuals from unpredictable social or interoceptive stimuli.

Finally, it is an unfortunate tragedy of life that some individuals feel so unsafe and uncomforted around others that they feel like the only option is to be as self-reliant as possible (A6), but in the DMM, this is recognized as a last resort \cite{CrittendenLandini2011}.\footnote{That is, individuals will not habituate to social isolation unless absolutely necessary for their safety. The DMM recognizes that it is only possible in adolescence when individuals can finally survive on their own.}

\subsection*{Type C}

Type C strategies arise as a result of persistent uncertainty during development that puts the individual in danger, and involve the strategic exaggeration of negative affect \cite{CrittendenLandini2011}.

 In infancy, Type C children are misled by caregivers as to their availability. Their resistant (C1) and ambivalent (C2) behaviors in the SSP are attempts to keep the caregiver engaged with them. Merely being in proximity to their caregiver was not enough to guarantee their availability -- only clear signals were guarantees, and their behaviors are organized around forcing those signals to be given. As a result, they suppress exteroceptive prediction errors once those errors are propagated to higher-level semantic regions that represent the caregiver's availability. However, lower-level sensory regions which generally have high fidelity regardless of the situation, such as the visual cortex, might not undergo the same suppression.

Despite their mistrust of exteroceptive cues, type C children tend to boost interoceptive prediction errors for two reasons:

\begin{itemize}
	\item There are few surer signs of availability than having interactions with their caregiver which promote an interoceptive response. Hence, their concept of availability resolves around the visceral feelings of interaction and not just mere proximity.
	\item During infancy, exploration is naturally highly rewarding, so infants are willing to tolerate some discomfort to learn about the world, so long as they have a consistent caregiver to return to when they need to find safety again. Yet, inconsistent caregiving does not inspire the same confidence in a type C individual, who will be less tolerant of discomfort. As a result, they are comfortable in a narrower range of interoceptive states, which results in hypervigilance to their own interoceptive state in general.
\end{itemize}

However, unlike type A individuals, type C individuals have not suffered from neglect or abuse consistently enough to force them to habituate to danger. If they feel endangered, then they will persevere, sometimes beyond reason, to avert the danger -- whether it is by resolving the threat or resolving the uncertainty.

Despite their low precision, type C individuals do still find exteroceptive observations relevant to their safety -- it is just that they obsessively seek that information to compensate for imprecise exteroceptive prediction errors, leading to frequent stress responses. They actively try to make exteroceptive precision errors as reliable as possible by paying focused attention to them, temporarily increasing their precision, and by using protest behaviors which make others act more predictably.

Upon reaching preschool age, the type C strategy can develop into overt displays of anger (C3) or vulnerability (C4), perhaps now also motivated by the knowledge that these overt displays will trigger a predictable response. The DMM even describes the possibility of alternating between exaggerating anger (odd-numbered type C strategies) to provoke a response and exaggerating vulnerability (even-numbered type C strategies) to disarm caregiver frustration \cite{Crittenden2016}. Even in preschool age, children are strategically adapting their attachment behaviors based on the caregiver's predicted mental state, all made possible by the maturation of the default mode network. 

Around school age, type C children can even learn to \textit{deceive} others into conceding to their anger (C5) or into rescuing them from fictitious dangers (C6). To deceive others, the child must be aware that others have a mental state separate from their own, which usually develops between age 4 and 6 \cite{Frith2003}, just before typical school age.

Type C strategies typically lead to fewer issues with interoception than type A strategies, since interoceptive inputs are not suppressed.\footnote{Although they might omit the \textit{expression} of anger or vulnerability during protest \cite{Crittenden2016}.} However, type C individuals struggle to trust exteroceptive evidence for their own safety, especially in their attachments, often leading to chronic stress. Over time, they might face difficulties integrating exteroception into their attachment concepts \cite{Crittenden2006}, and their most stable attachment concepts likely involve a fair deal of protest behavior which make exteroceptive cues more reliable (such as after successfully deceiving a caregiver into a clear response). Indeed, in AAI transcripts, type C individuals tend to give long and rambling responses which exaggerate their affect more than the contingencies of the overall situation itself \cite{CrittendenLandini2011}. When protest behaviors fail to work or even become self-sabotaging, they struggle to cope with the ambiguity and lack the attachment concepts required to distinguish between safety and danger. While type C individuals are able to mentalize in service of their strategy, they might struggle to mentalize over longer time periods without intervening action.

In the DMM, C5 and higher strategies (C5-8) involve a distrust of comfort \cite{CrittendenLandini2011}, which can make relationships challenging later in life. The C5 strategy is even considered to be equivalent to the \textit{dismissing} Ds2 classification in the Berkeley model. The use of deception to maintain availability precludes trust (or any substantial willingness to form a trusting relationship) in the long-term. As such, C5-8 individuals struggle to form attachment concepts not because of the lack of interoception, but because their strategies simply forbid them from forming and learning from trusting relationships. Mentalization is also jeopardized, since C5-8 strategies require a degree of disregard for the perspectives of others.

\section{Conclusion}
By taking the DMM -- an established theory of attachment -- and redefining its constructs based on predictive coding (specifically, active inference) and the EPIC model, our framework seamlessly merges those lines of research together. While the computational models on which our framework is based are supported by a growing body of evidence, our framework is still theoretical and its more specific predictions remain to be tested. Nonetheless, we hope to breathe new life into attachment research by placing it within the realm of modern computational neuroscience.

\clearpage
\bibliography{attachment-brain}

\end{document}